\documentclass[a4paper,11pt]{article}
\usepackage{jinstpub} 
\usepackage{lineno}

\title{\boldmath Measurement of gas properties for the ion-TPC of N$\nu$DEx experiment}

\author[a,b]{Tianyu Liang,}
\author[c,d]{Meiqiang Zhan,}
\author[a,b,1]{Hulin Wang\note{Corresponding author.},}
\author[c]{Xianglun Wei,}
\author[a,b]{Dongliang Zhang,}
\author[a,b]{Jun Liu,}
\author[c,2]{Chengui Lu\note{Corresponding author.},}
\author[c]{Qiang Hu,}
\author[c]{Yichen Yang,}
\author[a,b]{Chaosong Gao,}
\author[a,b]{Le Xiao,}
\author[a,b]{Xiangming Sun,}
\author[a,b]{Feng Liu,}
\author[c]{Chengxin Zhao,}
\author[c]{Hao Qiu,}
\author[a,b,3]{Kai Chen\note{Corresponding author.}}

\affiliation[a]{PLAC, Key Laboratory of Quark \& Lepton Physics (MOE), Central China Normal University, 152, Luoyu Rd. Wuhan, China}
\affiliation[b]{Hubei Provincial Engineering Research Center of Silicon Pixel Chip \& Detection Technology, 152, Luoyu Rd. Wuhan, China}
\affiliation[c]{Institute of Modern Physics, Chinese Academy of Sciences, 509, Nanchang Rd., Lanzhou, China}
\affiliation[d]{SiChuan University, 29, Wangjiang Rd., Chengdu, China}

\emailAdd{hulin.wang@ccnu.edu.cn}
\emailAdd{luchengui@impcas.ac.cn}
\emailAdd{chenkai@ccnu.edu.cn}

\abstract{
In the N$\nu$DEx collaboration, a high-pressure gas TPC is being developed to search for the neutrinoless double beta decay. The use of electronegative $\mathrm{^{82}SeF_{6}}$ gas mandates an ion-TPC. The reconstruction of $z$ coordinate is to be realized exploiting the feature of multiple species of charge carriers. As the initial stage of the development, we studied the properties of the $\mathrm{SF_{6}}$ gas, which is non-toxic and has similar molecular structure to $\mathrm{SeF_{6}}$. In the paper we present the measurement of drift velocities and mobilities of the majority and minority negative charge carriers found in $\mathrm{SF_{6}}$ at a pressure of 750 Torr, slightly higher than the local atmospheric pressure. The reduced fields range between 3.0 and 5.5 Td. It was performed using a laser beam to ionize the gas inside a small TPC, with a drift length of 3.7 cm. A customized charge sensitive amplifier was developed to read out the anode signals induced by the slowly drifting ions. The reconstruction of $z$ coordinate using the difference in the velocities of the two carriers was also demonstrated. 

}

\keywords{Double-beta decay detectors, Gaseous imaging and tracking detectors, Time projection Chambers (TPC), Charge transport and multiplication in gas}

\begin{document}
\maketitle
\flushbottom

\section{Introduction}
\label{sec:intro}

The search for neutrinoless double beta decay ($\mathrm{0\nu\beta\beta}$) is an idea way to probe the Majorana nature of the neutrinos. 
In addition, its discovery will prove that the lepton number is not conserved, and help to study the absolute masses of the neutrinos and the origin of their masses.
Recent or future experiments worldwide search for $\mathrm{0\nu\beta\beta}$ decay using diverse technologies, including the high-purity germanium detectors (GERDA~\cite{Agostini2018}, \textsc{majorana demonstrator}~\cite{MAJORANA2014}, LEGEND~\cite{legendc2021}, CDEX~\cite{cdex2017}), time-projection chambers (TPCs) with Xe (EXO-200~\cite{MAuger_2012}, nEXO~\cite{Adhikari_2022}, NEXT~\cite{next2023}, PandaX-III~\cite{PandaX2017}, LZ~\cite{lz2015}, DARWIN~\cite{Aalbers_2016}), liquid scintillators (KamLAND-Zen~\cite{PhysRevLett.130.051801}, SNO+~\cite{Albanese_2021}), cryogenic calorimeters (CUORE~\cite{ALDUINO20199}, CUPID~\cite{cupid2019}, CROSS~\cite{cross2020}, AMoRE~\cite{Lee_2020}), and tracking calorimeters (NEMO-3~\cite{PhysRevD.92.072011}, SuperNEMO~\cite{SuperNEMO2006}).

The N$\nu$DEx collaboration~\cite{nvdexcdr2023} proposed to search for $\mathrm{0\nu\beta\beta}$ decay of $\mathrm{^{82}Se}$ using high-pressure $\mathrm{^{82}SeF_{6}}$ gas TPC~\cite{Nygren_2018}.
The $\mathrm{SeF_{6}}$ has high electron affinity, and the electrons produced by the ionization process are quickly captured by the $\mathrm{SeF_{6}}$ molecules to form negative ions.
The negative ions are to be collected by the readout plane~\cite{10262350} consisting of CMOS integrated charge sensors named Topmetal-S~\cite{Yang_2024,Liang_2024}, without being amplified in the gas.
The main feature of Topmetal-S is its exposed metal on the top layer of the chip to sense the drifting charge carriers in the gas, hence integrating the functions of both charge sensor and readout application-specific integrated circuit in one chip.
With this design, low noise can be realized, which is important for direct sensing.
Both negative and position ions are envisaged to be detected by the charge sensors, while in the first phase of the experiment with 100 kg $\mathrm{^{82}SeF_{6}}$ at 10 bar~\cite{nvdexcdr2023}, only negative ions will be probed.

It is expected that in $\mathrm{SeF_{6}}$, the negative ions are predominately $\mathrm{SeF_{5}^{-}}$ and $\mathrm{SeF_{6}^{-}}$, like in $\mathrm{SF_{6}}$ which has similar molecular structure.
One advantage is that, by using the different drift velocities of the two charge carriers, the $z$ coordinate can be calculated without the need of start time, which is difficult to implement in N$\nu$DEx experiment.
The $\mathrm{SeF_{6}}$ is toxic and needs to be handled with care.
So in the first phase of the gas property study for N$\nu$DEx experiment, we begin by studying $\mathrm{SF_{6}}$ gas.
The properties of $\mathrm{SF_{6}}$ have been studied in low-pressure environment~\cite{Phan_2017} for the dark matter experiments.
In our study, we aim to observe and measure the properties of $\mathrm{SF_{5}^{-}}$ and $\mathrm{SF_{6}^{-}}$ around the atmospheric pressure, and to study the $z$ coordinate reconstruction.

The paper is structured as follows. 
The experimental apparatus and the method are described in Section~\ref{sec:exp}.
The results of the measurements, including the waveforms, the velocities and mobilities, are presented in Section~\ref{sec:meas}.
The reconstruction of $z$ coordinate using the different drift velocities of the two charge carriers is demonstrated in Section~\ref{sec:z}, followed by the conclusion in Section~\ref{sec:con}.

\section{Experimental apparatus and method}
\label{sec:exp}

\subsection{Device setup}
\label{subsec:dev}

\begin{figure}[htbp]
    \centering
    \includegraphics[width=0.96\textwidth]{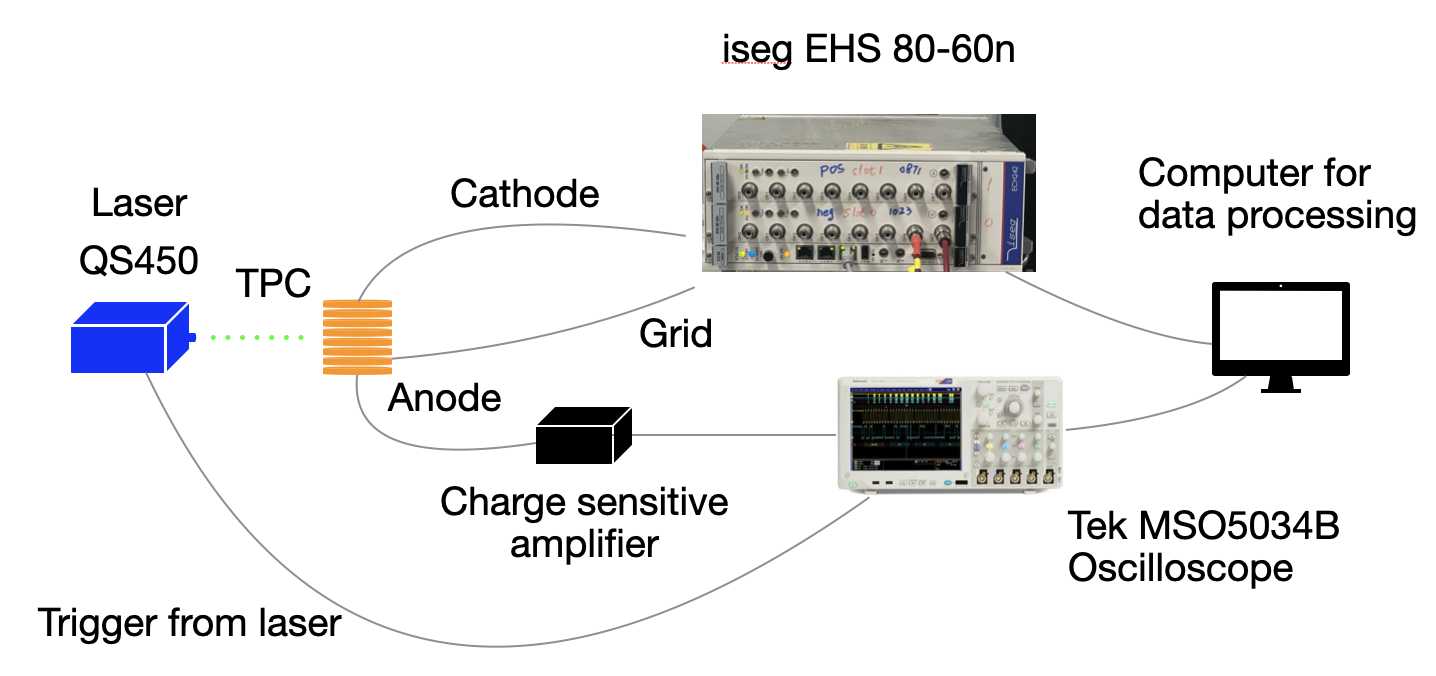}
    \caption{\label{fig:setup}The overall device setup.}
\end{figure}

The overall device setup for the measurement is shown in Figure~\ref{fig:setup}.
A Quantel Q-smart (450 mJ) laser is used to generate the pulsed laser beam with a wavelength of 266 nm.
It also sends a trigger signal which marks the start time of the event.
The laser beam ionizes the gas in the TPC, and the anode plane below the grid is used to collect the negative ions.
The anode plane is connected to a charge sensitive amplifier (CSA) which turns the current signal into the voltage signal.
Both the trigger signal and the signal from CSA are sent to a Tektronix MSO5034B oscilloscope for further analyses.

\begin{figure}[htbp]
    \centering
    \includegraphics[width=0.48\textwidth]{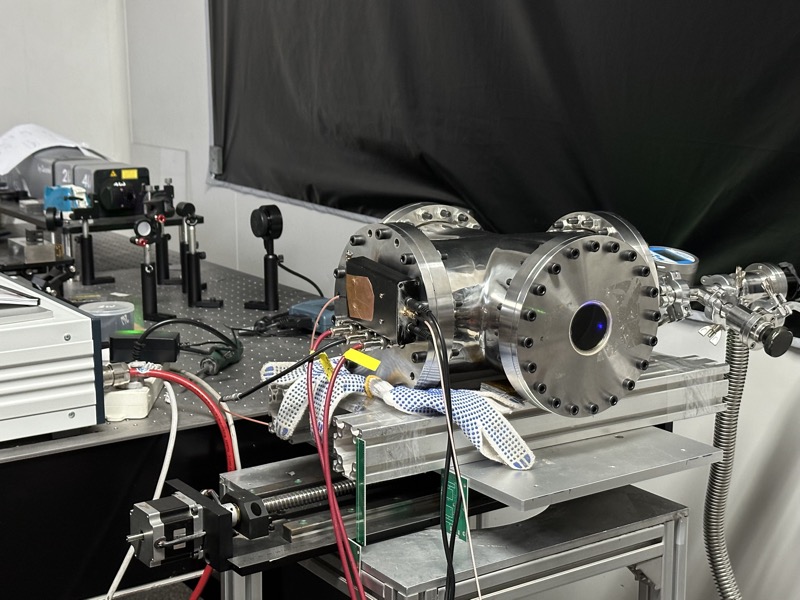}
    \includegraphics[width=0.48\textwidth]{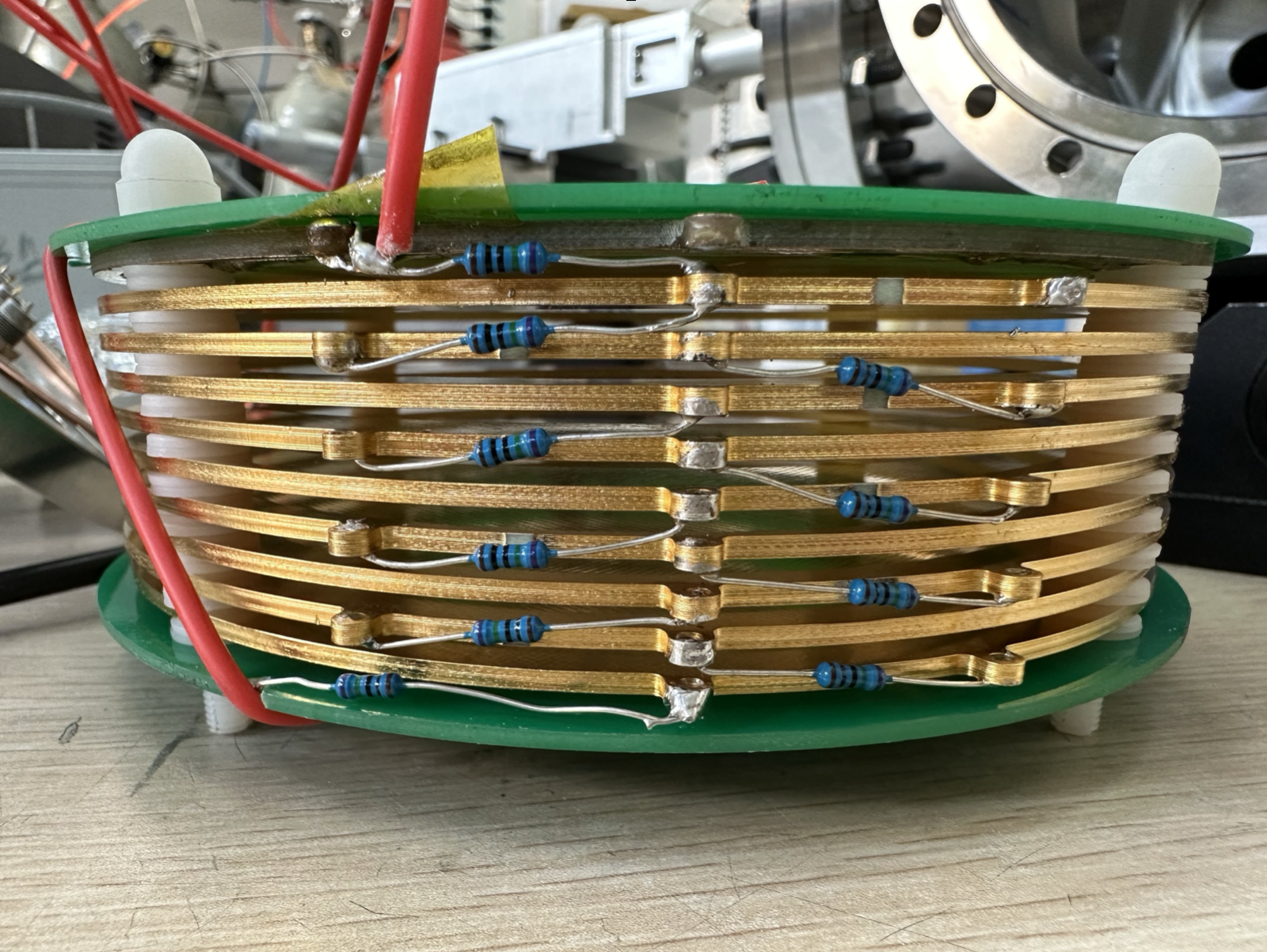}
    \caption{\label{fig:devicePhoto}The photo of the laser and the gas vessel (left) and the photo of the field cage (right).}
\end{figure}

The photo of the laser and the steel gas vessel is shown in Figure~\ref{fig:devicePhoto} (left).
The laser beam enters the field cage of the TPC through a quartz window in the gas vessel.
The photo of the field cage inside the gas vessel is shown in Figure~\ref{fig:devicePhoto} (right).
It has a length of 40 mm, and the maximum drift distance (between cathode and grid) is 38 mm.
The grid is made of wires of 50 $\mu$m diameter, and there is a 2 mm gap between the grid and the anode.
The field rings are made of 2 mm wide copper strips at a pitch of 4 mm.
They are connected by the 10 $\mathrm{M\Omega}$ resistors.

\subsection{The CSA}
\label{subsec:csa}

\begin{figure}[htbp]
    \centering
    \includegraphics[width=0.63\linewidth]{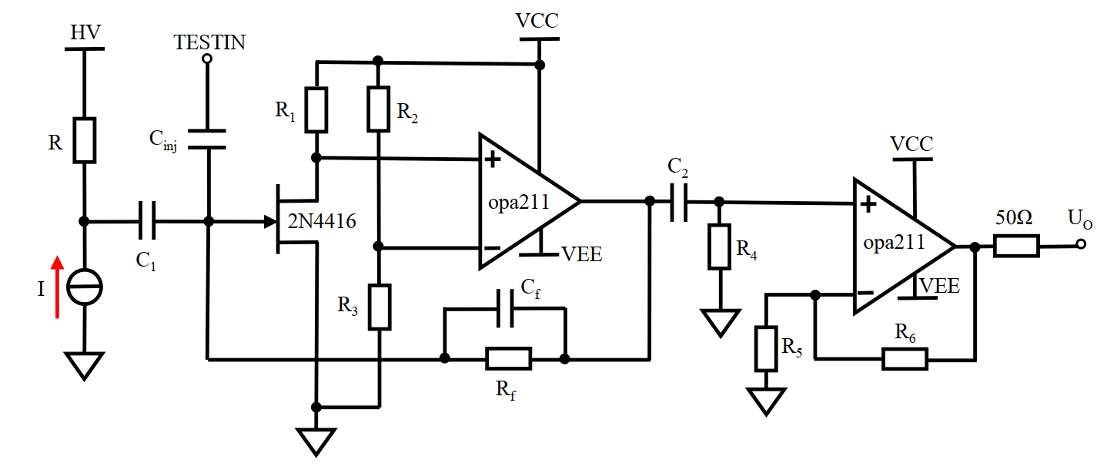}
    \includegraphics[width=0.33\linewidth]{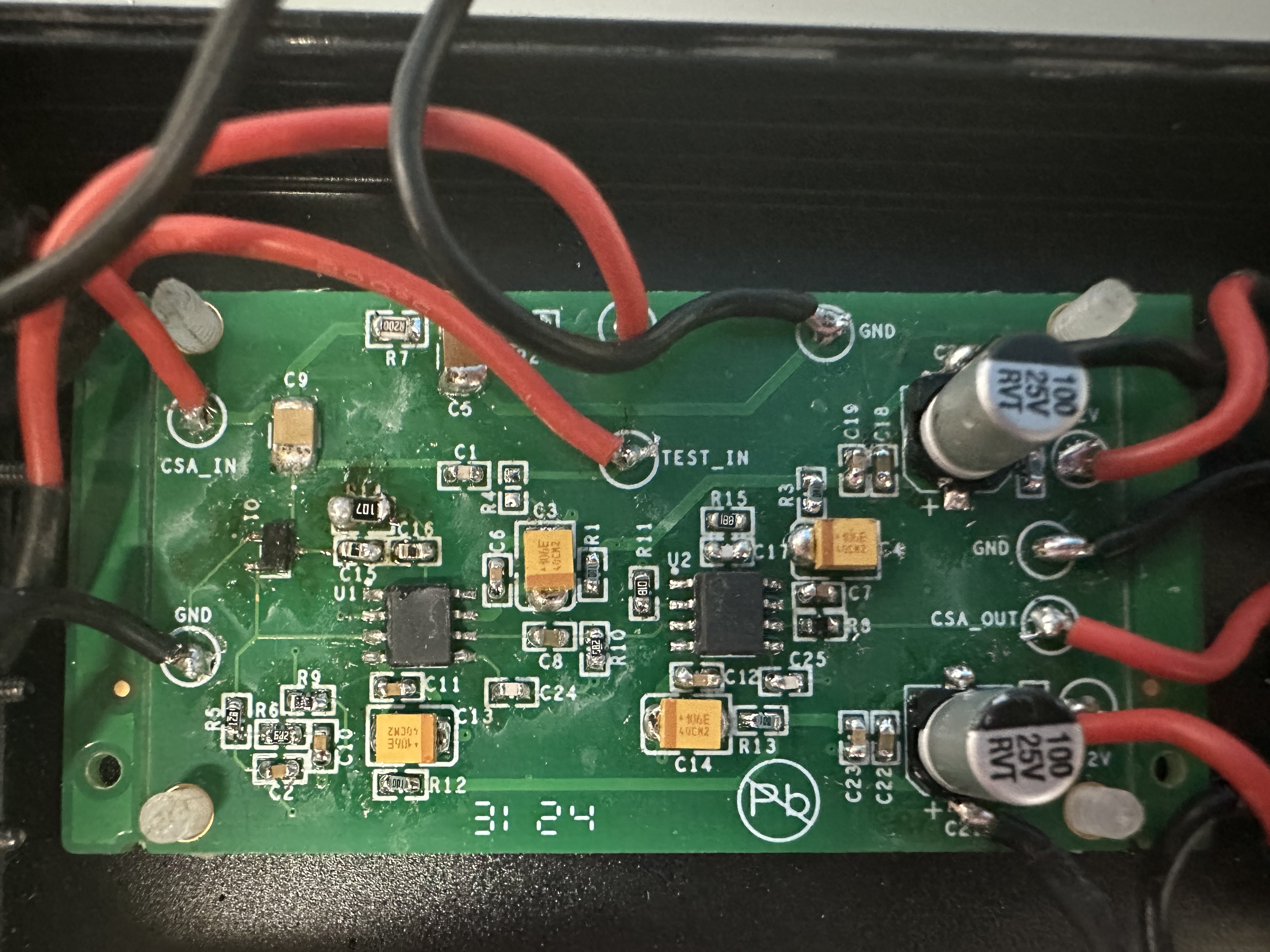}
    \caption{\label{fig:csa} The structure (left) and photo (right) of the CSA used in the measurement.}
\end{figure}

To detect the signal induced by the slowly drifting ions, a CSA has been specifically developed for the measurement, with its structure and photo shown in Figure~\ref{fig:csa}. 

It has two amplification stages.
The current signal from the anode plane passes through the blocking capacitor $\mathrm{C_{1}}$, and then enters into the first amplification stage,
which is composed of a transistor and an OPA211~\cite{opamanual} amplifier chip.
A JFET 2N4416 is used for its small gate leakage current and high transconductance.
The resistors R$_{1}$, R$_{2}$, R$_{3}$ are used to adjust the quiescent point of the JFET to make it work in the amplification regime.
The decay time constant of the CSA is determined by the product of the feedback resistor $\mathrm{R_{f}}$ and feedback capacitor $\mathrm{C_{f}}$.
The $\mathrm{C_{2}}$ and $\mathrm{R_{4}}$ form a high pass filter.
The output of the first stage is filtered and then enters into the second amplification stage with another OPA211 chip, with the gain determined by $\mathrm{R_{6}}$/$\mathrm{R_{5}}$. 
The output impedance is 50 $\Omega$.
A test pulse could be injected into the TESTIN pin to calibrate the CSA with the injection capacitance $\mathrm{C_{inj}}$.

The values of $\mathrm{C_{f}}$ and $\mathrm{R_{f}}$ were adjusted to achieve desired decay time constant and noise performance of the CSA. 
In the subsequent measurements, 50 fF and 1 $\mathrm{G\Omega}$ were adopted for $\mathrm{C_{f}}$ and $\mathrm{R_{f}}$, respectively.
The main reason for the rather short 50 $\mu$s decay time constant was to avoid the saturation of the CSA during the measurement,
which will be improved to allow longer decay time in the future.

\section{Results of measurement}
\label{sec:meas}

\subsection{Waveforms of $\mathrm{SF_{6}}$}
\label{subsec:wav}

The laser beam, with a diameter of about 1 mm, enters the field cage through a 2 mm wide gap below the cathode.
The laser beam is parallel to the cathode, and the distance between the beam center and grid is 37 mm, which defines the drift length of the measurement.
The trigger signal from the laser gives the start time of the event.
The gas pressure is 750 Torr, slightly higher than the local atmospheric pressure of 630 Torr.
The measurement was carried out at room temperature of about 20$^{\circ}$C.
The electric field strength of the drift region varies between 735 V/cm and 1351 V/cm.
The electric field strength between grid and anode is fixed to be 5000 V/cm.

The waveforms of the voltage signals from the CSA are acquired by the oscilloscope. 
The current signal is calculated from the voltage signal using the equation~\cite{Phan_2017}:
\begin{equation}
\label{eq:i}
I(t) \propto \frac{dV}{dt} - (-\frac{V}{\tau})
\end{equation}
where $\tau$ is the decay time constant of the CSA.
The current waveforms are then smoothed with a Butterworth filter to suppress high-frequency noise.

\begin{figure}[htbp]
    \centering
    \includegraphics[width=0.48\linewidth]{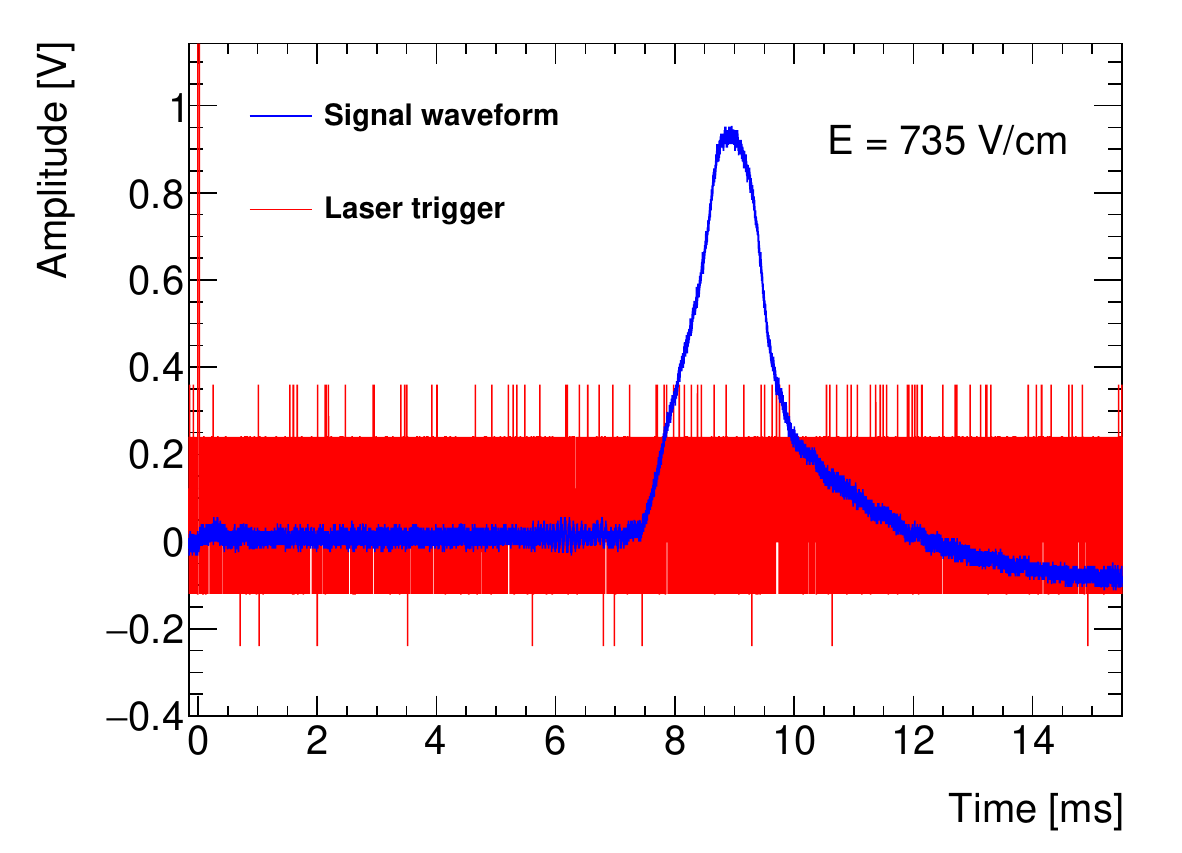}
    \includegraphics[width=0.48\linewidth]{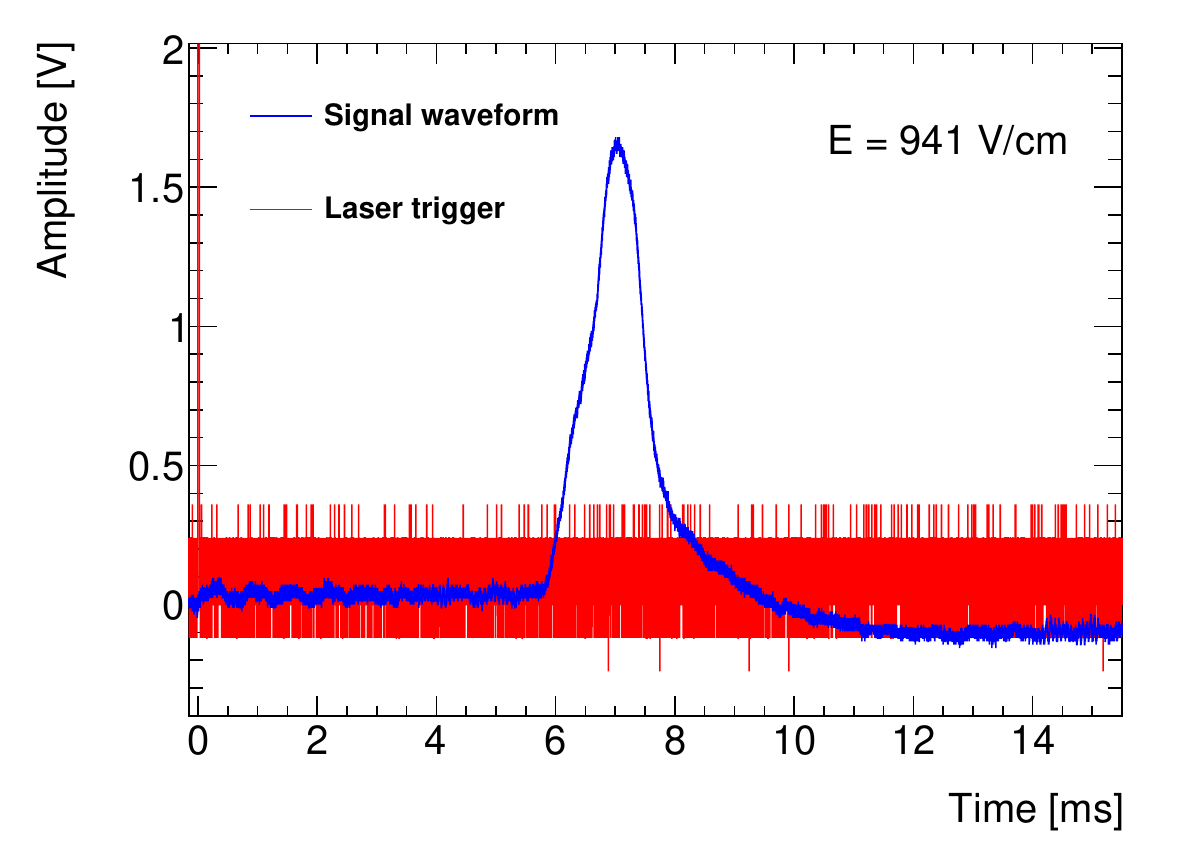}
    \includegraphics[width=0.48\linewidth]{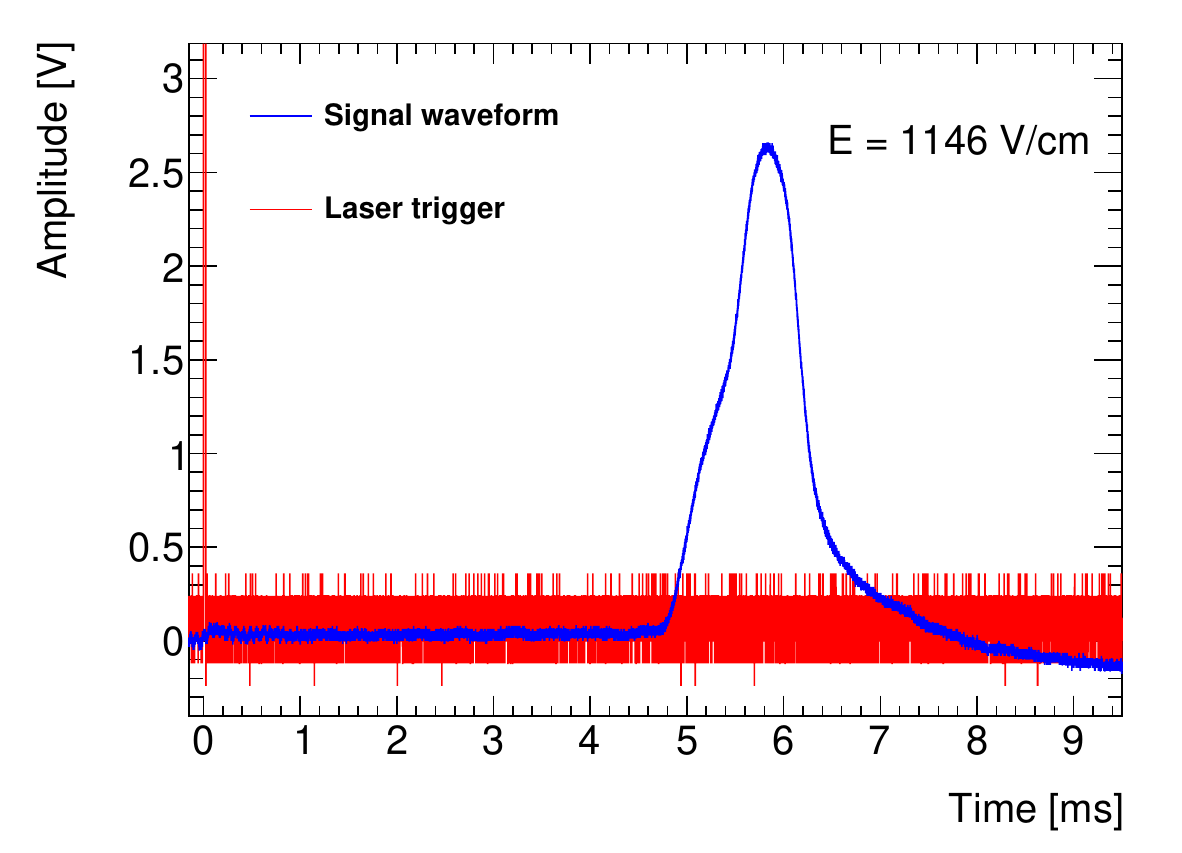}
    \includegraphics[width=0.48\linewidth]{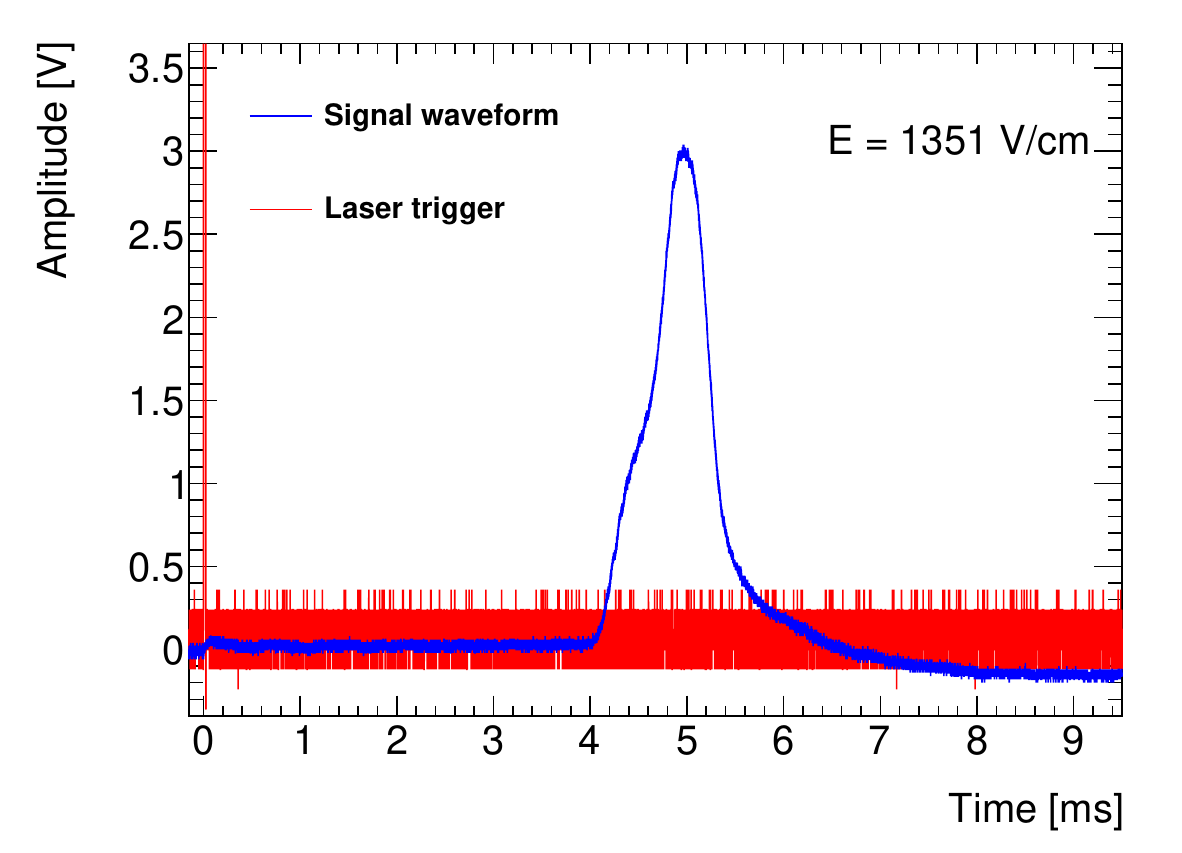}
    \caption{\label{fig:wavvol}The typical waveforms of the voltage signals, for four drift fields between 735 V/cm and 1351 V/cm. The trigger signals from the laser are also shown.}
\end{figure}

\begin{figure}[htbp]
    \centering
    \includegraphics[width=0.48\linewidth]{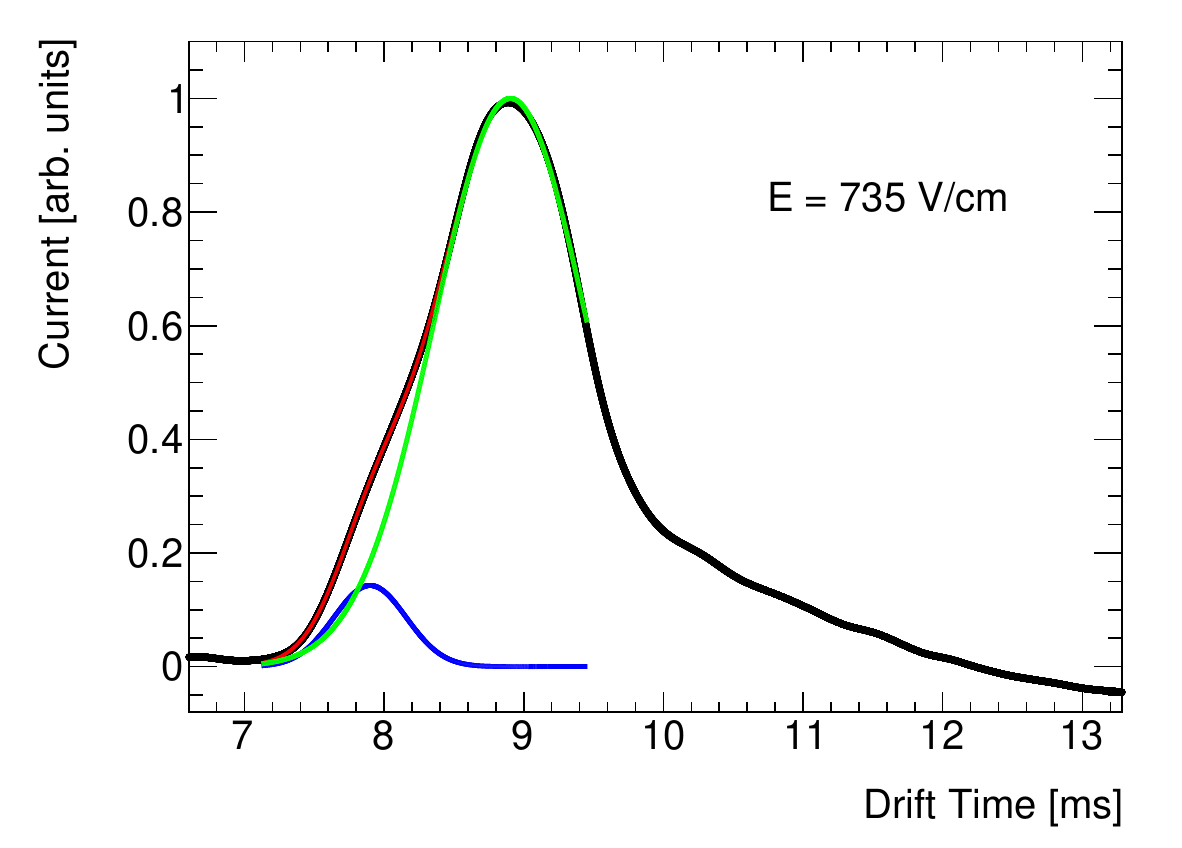}
    \includegraphics[width=0.48\linewidth]{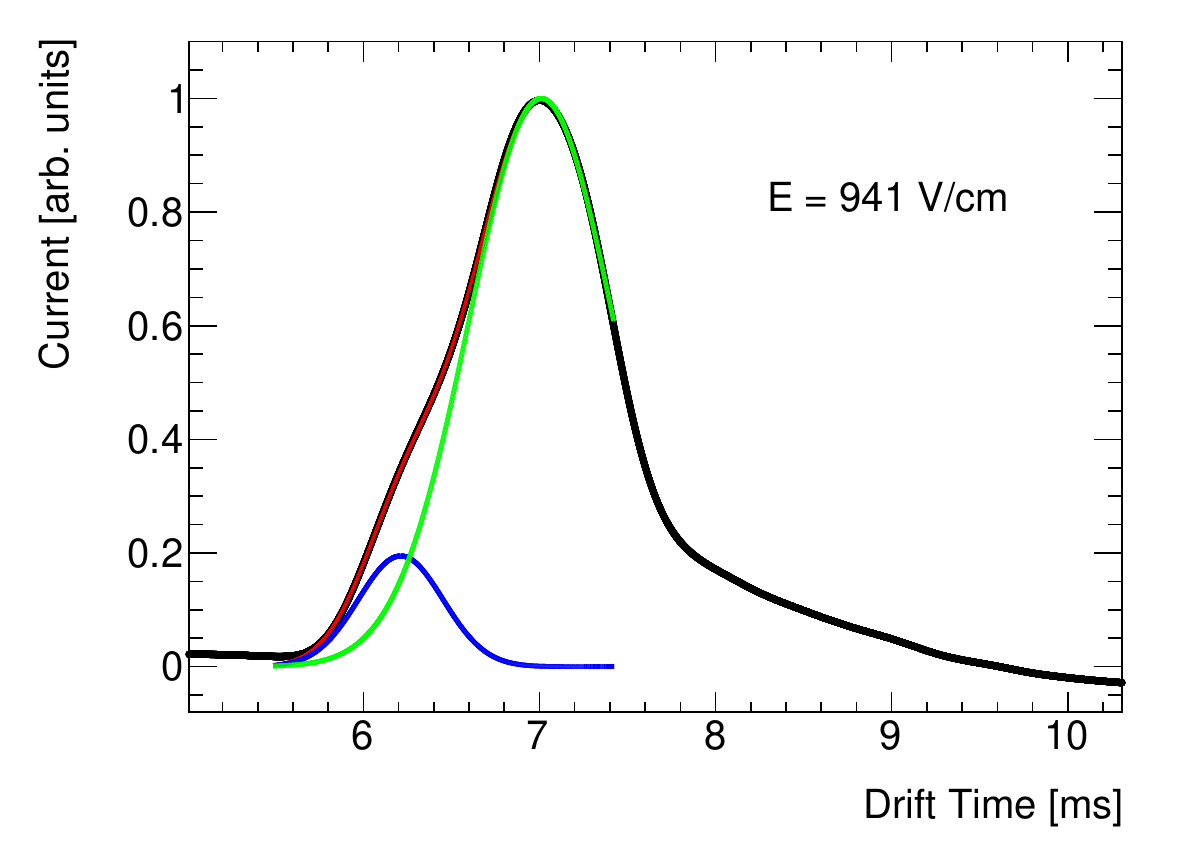}
    \includegraphics[width=0.48\linewidth]{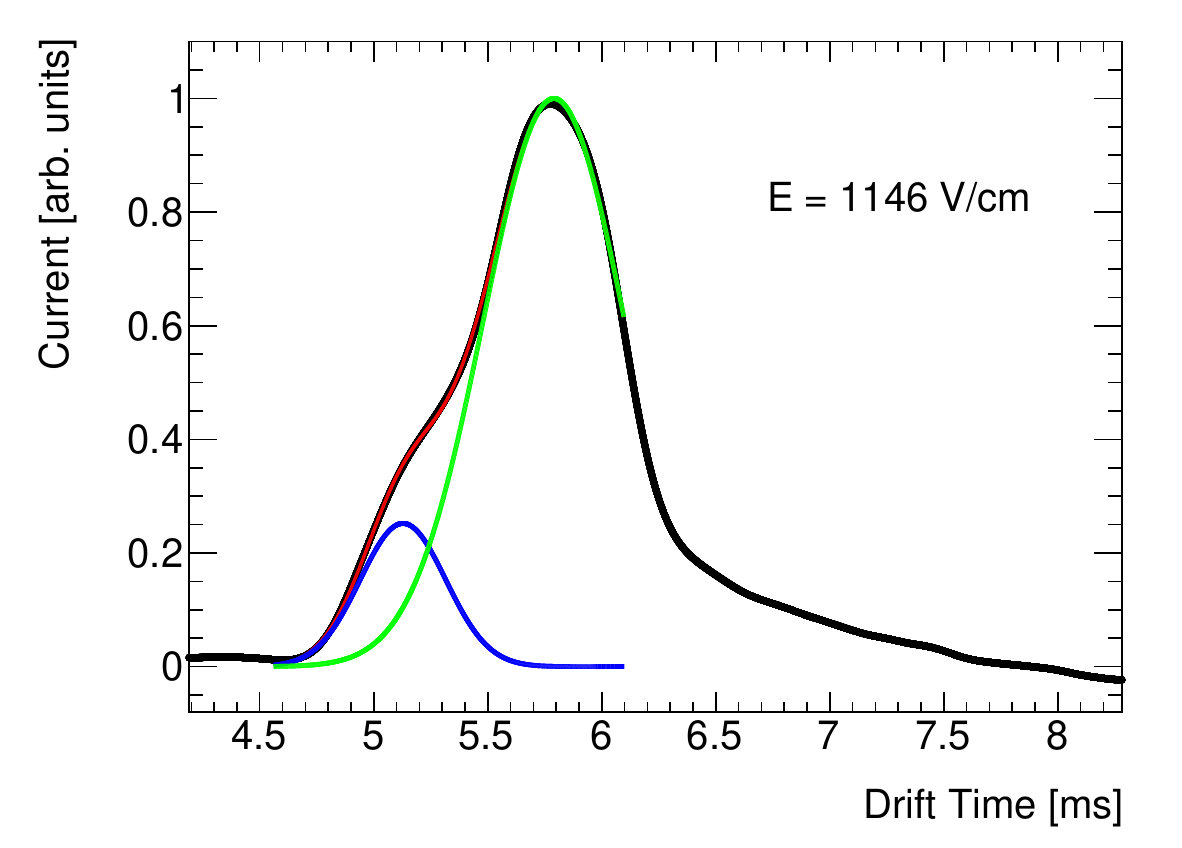}
    \includegraphics[width=0.48\linewidth]{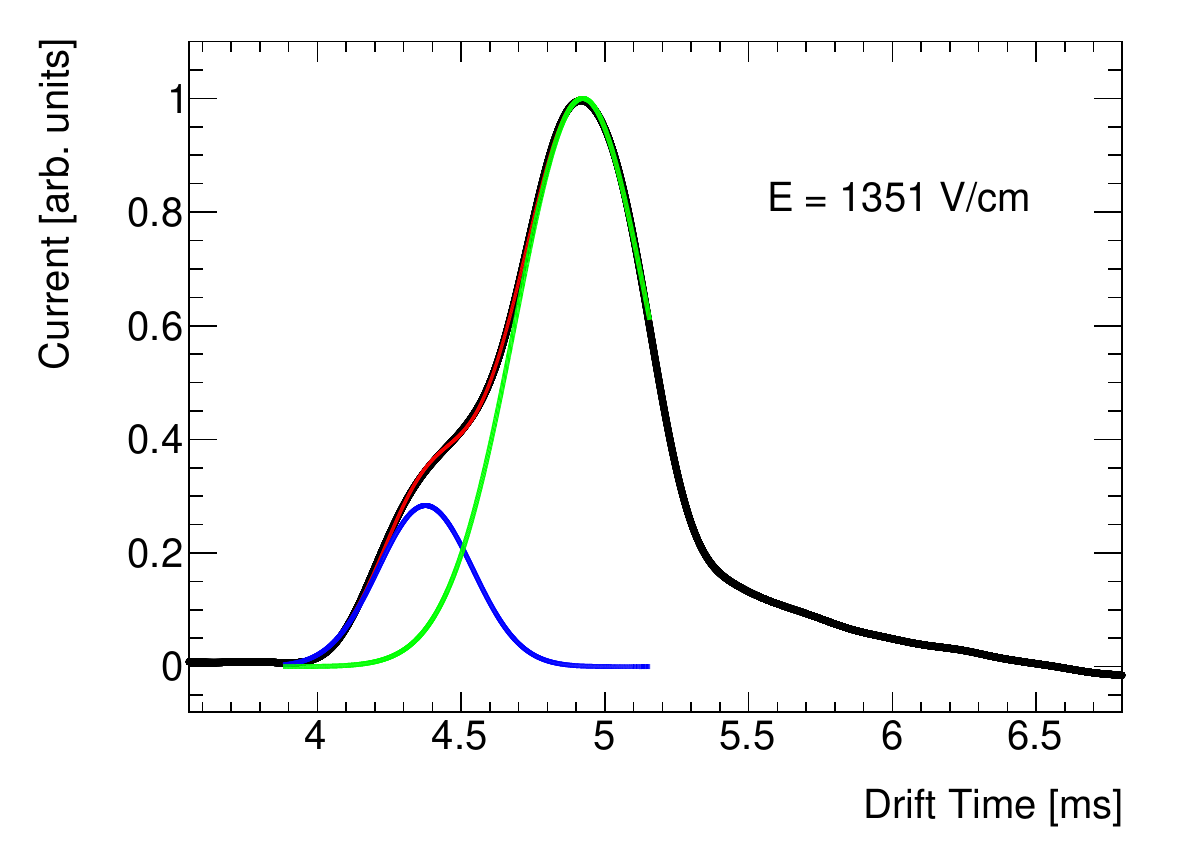}
    \caption{\label{fig:wavcur}The typical waveforms of the current signals, converted from the voltage waveforms using Eq.~\ref{eq:i}, for four drift fields between 735 V/cm and 1351 V/cm.
    The amplitude of the majority peak is normalized to unity. 
    The double-Gaussian fit to each waveform is also shown, to extract the separate drift times of the minority and majority charge carriers.
    }
\end{figure}

Figure~\ref{fig:wavvol} and~\ref{fig:wavcur} show the typical waveforms of the voltage signals and the resulting current signals, respectively. 
The trigger signals from the laser are also shown in the voltage waveforms.
Due to the small decay time of CSA, the waveform of current signal is similar to the corresponding waveform of voltage signal. 
For the current waveforms, the amplitudes of the majority peaks are normalized to unit.

Due to the small drift length and finite laser beam width, the waveforms of two charge carriers are not quite separated from each other.
But it can indeed be seen that there are at least two charge carriers in the current waveform, and the minority peak becomes more obvious for higher drift field.
The minority charge carrier and the faster one is postulated to be $\mathrm{SF_{5}^{-}}$, while the majority charge carrier and the slower one is postulated to be $\mathrm{SF_{6}^{-}}$.
There is also long tail on the right side of the majority peak.
This could be due to $\mathrm{SF_{5}^{-}}(\mathrm{SF_{6}})_{n}$ and $\mathrm{SF_{6}^{-}}(\mathrm{SF_{6}})_{n}$, as well as $\mathrm{SF_{6}^{-}}(\mathrm{H_{2}O})_{n}$,
which drift at slower speed than the $\mathrm{SF_{6}^{-}}$.

The ratio of the amplitude of the minority carrier to that of the majority carrier increases from 0.13 to 0.28, as the drift field increases from 735 V/cm to 1351 V/cm.
These are larger than the values reported in the measurement in low pressure~\cite{Phan_2017}.
We suspect that it is mainly due to the different signal generation mechanisms, i.e. we use laser beam to ionize the gas.
The cross section of $\mathrm{SF_{5}^{-}}$ has strong dependence on the electron energy, and could be enhanced in our case.
The relative amplitude between $\mathrm{SF_{5}^{-}}$ and $\mathrm{SF_{6}^{-}}$ in different conditions will be further studied in the next step. 

\subsection{Drift velocities and mobilities}
\label{subsec:vec}

\begin{figure}[htbp]
    \centering
    \includegraphics[width=0.98\linewidth]{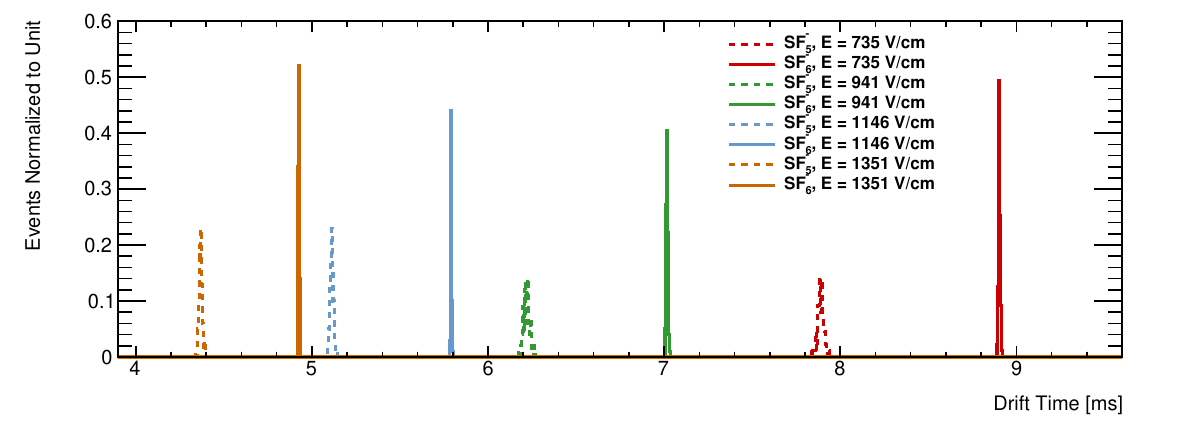}
    \caption{\label{fig:dtime}The distributions of drift times for $\mathrm{SF_{5}^{-}}$ and $\mathrm{SF_{6}^{-}}$, for four drift fields between 735 V/cm and 1351 V/cm.}
\end{figure}

The drift times of $\mathrm{SF_{5}^{-}}$ and $\mathrm{SF_{6}^{-}}$ are extracted by fitting a double-Gaussian function to the current waveform to extract their separate contributions,
and the two $\mu$ values are used as their drift times.
Examples of the double-Gaussian fit are shown in Figure~\ref{fig:wavcur}, for four different drift fields.
Due to the long tail at the right side of majority peak as explained in Section~\ref{subsec:csa}, the range of the fit is only up to about one standard deviation above the majority peak.
The distributions of drift times for $\mathrm{SF_{5}^{-}}$ and $\mathrm{SF_{6}^{-}}$ are shown in Figure~\ref{fig:dtime}, for four different drift fields.

\begin{figure}[htbp]
    \centering
    \includegraphics[width=0.48\linewidth]{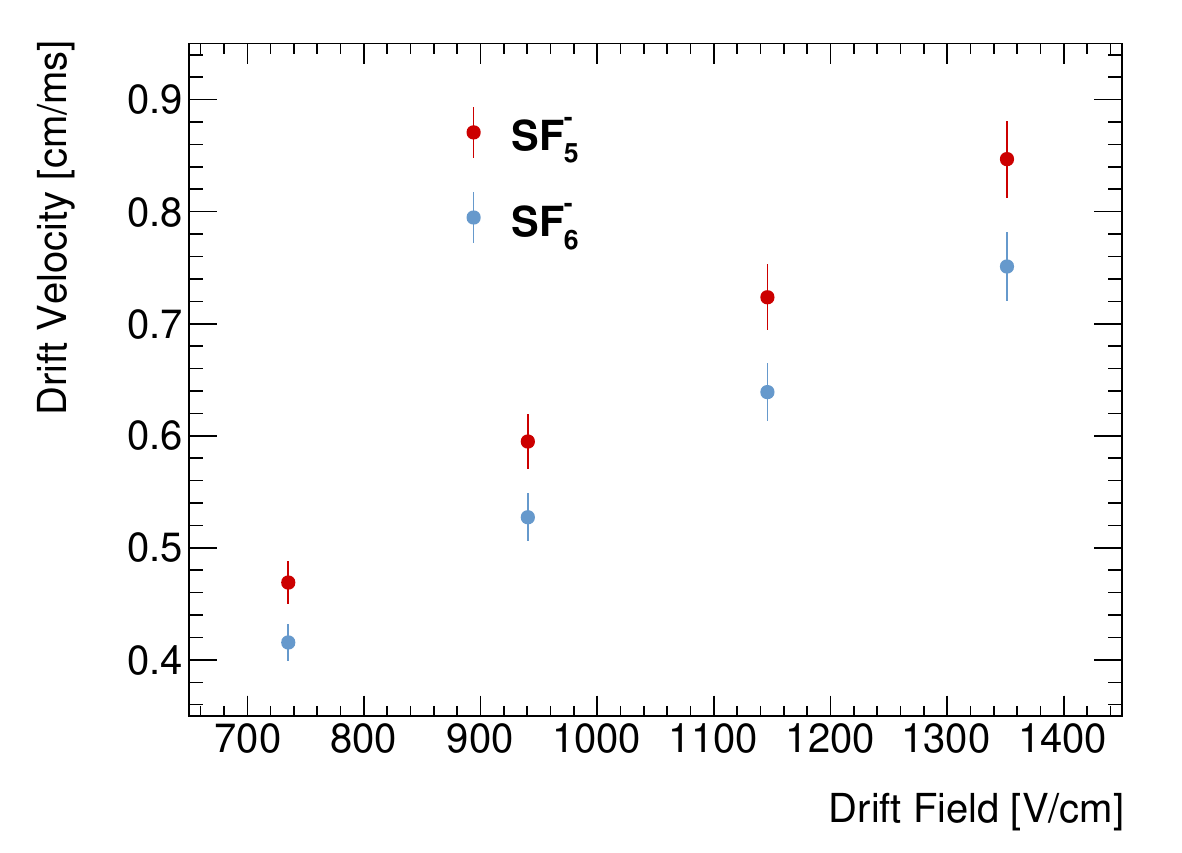}
    \includegraphics[width=0.48\linewidth]{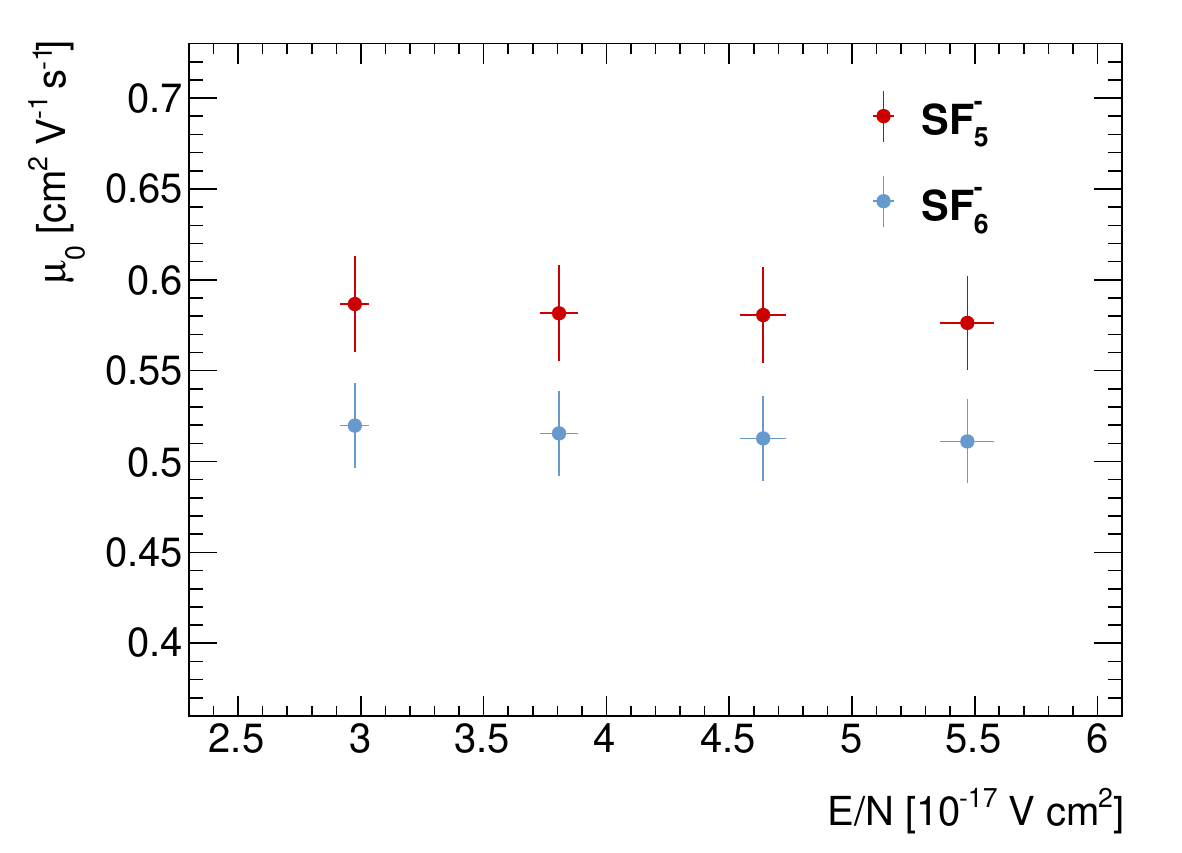}
    \caption{\label{fig:vecmob}The drift velocities (left) and reduced mobilities (right) of $\mathrm{SF_{5}^{-}}$ and $\mathrm{SF_{6}^{-}}$.}
\end{figure}

The drift velocity $v_{\mathrm{d}}$ is then calculated as the ratio between the 37 mm drift distance and the drift time.
A total 1.5 mm systematic uncertainty is considered on the drift length, including a $\sim$ 1 mm uncertainty for the position and width of the laser beam, and $\sim$ 0.5 mm uncertainty for the impact of 2 mm gap between grid and anode at 5000 V/cm. 
Figure~\ref{fig:vecmob} (left) shows the drift velocities of $\mathrm{SF_{5}^{-}}$ and $\mathrm{SF_{6}^{-}}$ as a function of drift field.

The reduced mobility $\mu_{0}$ is also calculated, defined as:
\begin{equation}
\label{eq:mobi}
\mu_{0} = \frac{v_{\mathrm{d}}}{E}\frac{N}{N_{0}}
\end{equation}
where $N_{\mathrm{0}} = 2.687 \times 10^{19}$ $\mathrm{cm^{-3}}$ is the gas density at STP ($0^{\circ}$C and 760 Torr).
A conservative 2$\%$ uncertainty is considered on the pressure and temperature.
Figure~\ref{fig:vecmob} (right) shows the reduced mobilities of 
$\mathrm{SF_{5}^{-}}$ and $\mathrm{SF_{6}^{-}}$ as a function of the 
reduced field $E/N$ in Townsend units (1 Td = $10^{-17}$ $\mathrm{V cm^{2}}$).

The reduced fields range between 3.0 and 5.5 Td.
The reduced mobilities of $\mathrm{SF_{5}^{-}}$ vary between 0.576 $\pm$ 0.026 $\mathrm{cm^{2}V^{-1}s^{-1}}$ and 0.587 $\pm$ 0.027 $\mathrm{cm^{2}V^{-1}s^{-1}}$,
while the reduced mobilities of $\mathrm{SF_{6}^{-}}$ vary between 0.511 $\pm$ 0.023 $\mathrm{cm^{2}V^{-1}s^{-1}}$ and 0.520 $\pm$ 0.024 $\mathrm{cm^{2}V^{-1}s^{-1}}$.
The values are consistent with what was reported in Ref.~\cite{Phan_2017} under the uncertainties.

\section{Reconstruction of $z$ coordinate}
\label{sec:z}

Without using the start time, the $z$ coordinate could be reconstructed by using the difference in the arrival times of $\mathrm{SF_{5}^{-}}$ and $\mathrm{SF_{6}^{-}}$, and their pre-calibrated drift velocities.
The $z$ coordinate reconstruction is demonstrated using the same datasets used in Section~\ref{sec:meas}, following the equation:
\begin{equation}
\label{eq:z}
z = \frac{v_{\mathrm{SF_{5}^{-}}} \cdot v_{\mathrm{SF_{6}^{-}}}}{v_{\mathrm{SF_{5}^{-}}}- v_{\mathrm{SF_{6}^{-}}}} \cdot \Delta t
\end{equation}
where $\Delta t$ is the difference in the arrival times of $\mathrm{SF_{5}^{-}}$ and $\mathrm{SF_{6}^{-}}$ in each event, while $v_{\mathrm{SF_{5}^{-}}}$ and $v_{\mathrm{SF_{6}^{-}}}$ are the measured drift velocities in Section~\ref{subsec:vec}.

\begin{figure}[htbp]
    \centering
    \includegraphics[width=0.7\linewidth]{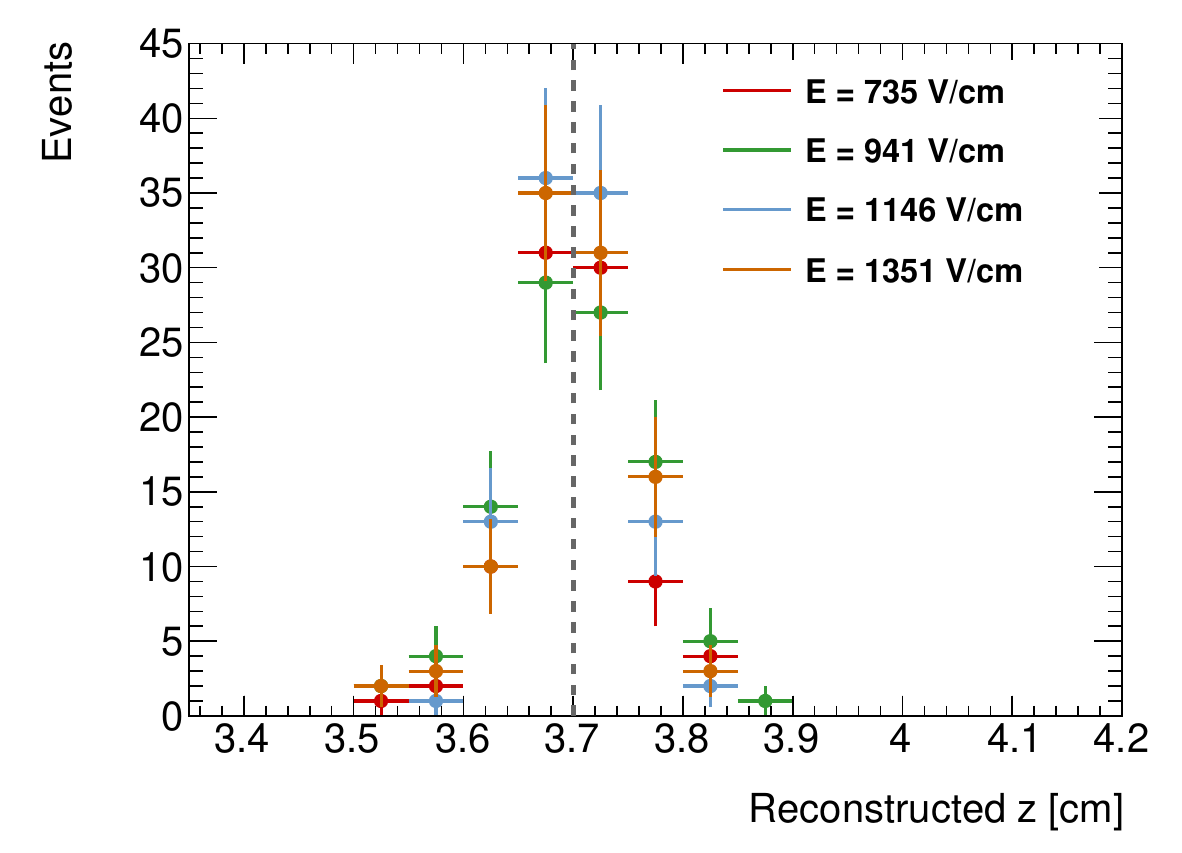}
    \caption{\label{fig:z}The distributions of the reconstructed z coordinate of the laser beam, for four drift fields between 735 V/cm and 1351 V/cm.}
\end{figure}

Figure~\ref{fig:z} shows the distributions of the reconstructed z coordinate of the laser beam, for four drift fields between 735 V/cm and 1351 V/cm.
The distributions center at 3.7 cm as expected, with the standard deviations varying from 0.046 to 0.067 cm.

\section{Conclusion}
\label{sec:con}

For the first phase of the gas property study for the N$\nu$DEx experiment using high-pressure $\mathrm{^{82}SeF_{6}}$ gas TPC to search for $\mathrm{0\nu\beta\beta}$ decay, 
a structurally similar $\mathrm{SF_{6}}$ gas was examined, at 750 Torr and room temperature, with the drift fields ranging from 735 V/cm to 1351 V/cm.
The corresponding reduced fields range from 3.0 to 5.5 Td.

The minority charge carrier ${\mathrm{SF_{5}^{-}}}$ and the majority charge carrier ${\mathrm{SF_{6}^{-}}}$ were observed in the waveforms, 
and their drift velocities and mobilities were measured, using a small TPC and a custom-developed CSA.
The signals were generated by ionizing the gas using a laser beam.

The reconstruction of $z$ coordinate using the difference in the arrival times of the two negative ion species were demonstrated.

The work in the paper lays the groundwork for future studies on the properties of $\mathrm{^{82}SeF_{6}}$ gas.
Measurements will be improved, and be extended to higher pressures close to the value in the N$\nu$DEx experiment, to positive ions, and to other gas properties including diffusion, W value, and the Fano factors.

\acknowledgments

This work was supported in part by the National Natural Science Foundation of China under Grant 12105110,
and in part by the National Key Research and Development Program of China under Grant 2022YFA1604703, 2021YFA1601300.


\bibliographystyle{JHEP}
\bibliography{NvDExGas2024.bib}

\providecommand{\href}[2]{#2}\begingroup\raggedright\begin{thebibliography}{10}

\bibitem{Agostini2018}
M.~Agostini, A.M.~Bakalyarov, M.~Balata, I.~Barabanov, L.~Baudis, C.~Bauer
  et~al., \emph{Upgrade for phase ii of the gerda experiment},
  \href{https://doi.org/10.1140/epjc/s10052-018-5812-2}{\emph{The European
  Physical Journal C} {\bfseries 78} (2018) 388}.

\bibitem{MAJORANA2014}
N.~Abgrall, E.~Aguayo, F.T.~Avignone~III, A.S.~Barabash, F.E.~Bertrand,
  M.~Boswell et~al., \emph{The majorana demonstrator neutrinoless double-beta
  decay experiment},
  \href{https://doi.org/https://doi.org/10.1155/2014/365432}{\emph{Advances in
  High Energy Physics} {\bfseries 2014} (2014) 365432}
  [\href{https://arxiv.org/abs/https://onlinelibrary.wiley.com/doi/pdf/10.1155/2014/365432}{{\ttfamily
  https://onlinelibrary.wiley.com/doi/pdf/10.1155/2014/365432}}].

\bibitem{legendc2021}
L.~Collaboration, N.~Abgrall, I.~Abt, M.~Agostini, A.~Alexander, C.~Andreoiu
  et~al., \emph{Legend-1000 preconceptual design report},  2021.

\bibitem{cdex2017}
L.~Wang, Q.~Yue, K.~Kang, J.~Cheng, Y.~Li, T.H.~Wong et~al., \emph{First
  results on 76ge neutrinoless double beta decay from cdex-1 experiment},
  \href{https://doi.org/10.1007/s11433-017-9038-4}{\emph{Science China Physics,
  Mechanics \& Astronomy} {\bfseries 60} (2017) 071011}.

\bibitem{MAuger_2012}
M.~Auger, D.J.~Auty, P.S.~Barbeau, L.~Bartoszek, E.~Baussan, E.~Beauchamp
  et~al., \emph{The exo-200 detector, part i: detector design and
  construction},
  \href{https://doi.org/10.1088/1748-0221/7/05/P05010}{\emph{Journal of
  Instrumentation} {\bfseries 7} (2012) P05010}.

\bibitem{Adhikari_2022}
G.~Adhikari, S.A.~Kharusi, E.~Angelico, G.~Anton, I.J.~Arnquist, I.~Badhrees
  et~al., \emph{nexo: neutrinoless double beta decay search beyond 1028 year
  half-life sensitivity},
  \href{https://doi.org/10.1088/1361-6471/ac3631}{\emph{Journal of Physics G:
  Nuclear and Particle Physics} {\bfseries 49} (2021) 015104}.

\bibitem{next2023}
P.~Novella, M.~Sorel, A.~Us{\'o}n, C.~Adams, H.~Almaz{\'a}n, V.~{\'A}lvarez
  et~al., \emph{Demonstration of neutrinoless double beta decay searches in
  gaseous xenon with next},
  \href{https://doi.org/10.1007/JHEP09(2023)190}{\emph{Journal of High Energy
  Physics} {\bfseries 2023} (2023) 190}.

\bibitem{PandaX2017}
X.~Chen, C.~Fu, J.~Galan, K.~Giboni, F.~Giuliani, L.~Gu et~al.,
  \emph{Pandax-iii: Searching for neutrinoless double beta decay with high
  pressure 136xe gas time projection chambers},
  \href{https://doi.org/10.1007/s11433-017-9028-0}{\emph{Science China Physics,
  Mechanics \& Astronomy} {\bfseries 60} (2017) 061011}.

\bibitem{lz2015}
T.L.~Collaboration, D.S.~Akerib, C.W.~Akerlof, D.Y.~Akimov, S.K.~Alsum,
  H.M.~Araújo et~al., \emph{Lux-zeplin (lz) conceptual design report},  2015.

\bibitem{Aalbers_2016}
J.~Aalbers, F.~Agostini, M.~Alfonsi, F.~Amaro, C.~Amsler, E.~Aprile et~al.,
  \emph{Darwin: towards the ultimate dark matter detector},
  \href{https://doi.org/10.1088/1475-7516/2016/11/017}{\emph{Journal of
  Cosmology and Astroparticle Physics} {\bfseries 2016} (2016) 017}.

\bibitem{PhysRevLett.130.051801}
{\scshape KamLAND-Zen Collaboration} collaboration, \emph{Search for the
  majorana nature of neutrinos in the inverted mass ordering region with
  kamland-zen},
  \href{https://doi.org/10.1103/PhysRevLett.130.051801}{\emph{Phys. Rev. Lett.}
  {\bfseries 130} (2023) 051801}.

\bibitem{Albanese_2021}
T.S.~collaboration, V.~Albanese, R.~Alves, M.~Anderson, S.~Andringa, L.~Anselmo
  et~al., \emph{The sno+ experiment},
  \href{https://doi.org/10.1088/1748-0221/16/08/P08059}{\emph{Journal of
  Instrumentation} {\bfseries 16} (2021) P08059}.

\bibitem{ALDUINO20199}
C.~Alduino, F.~Alessandria, M.~Balata, D.~Biare, M.~Biassoni, C.~Bucci et~al.,
  \emph{The cuore cryostat: An infrastructure for rare event searches at
  millikelvin temperatures},
  \href{https://doi.org/https://doi.org/10.1016/j.cryogenics.2019.06.011}{\emph{Cryogenics}
  {\bfseries 102} (2019) 9}.

\bibitem{cupid2019}
T.C.I.~Group, \emph{Cupid pre-cdr},  2019.

\bibitem{cross2020}
I.C.~Bandac, A.S.~Barabash, L.~Berg{\'e}, M.~Bri{\`e}re, C.~Bourgeois,
  P.~Carniti et~al., \emph{The $0\nu 2\beta$-decay cross experiment:
  preliminary results and prospects},
  \href{https://doi.org/10.1007/JHEP01(2020)018}{\emph{Journal of High Energy
  Physics} {\bfseries 2020} (2020) 18}.

\bibitem{Lee_2020}
M.~Lee, \emph{Amore: a search for neutrinoless double-beta decay of 100mo using
  low-temperature molybdenum-containing crystal detectors},
  \href{https://doi.org/10.1088/1748-0221/15/08/C08010}{\emph{Journal of
  Instrumentation} {\bfseries 15} (2020) C08010}.

\bibitem{PhysRevD.92.072011}
{\scshape NEMO-3 Collaboration} collaboration, \emph{Results of the search for
  neutrinoless double-$\ensuremath{\beta}$ decay in $^{100}\mathrm{Mo}$ with
  the nemo-3 experiment},
  \href{https://doi.org/10.1103/PhysRevD.92.072011}{\emph{Phys. Rev. D}
  {\bfseries 92} (2015) 072011}.

\bibitem{SuperNEMO2006}
F.~Piquemal, \emph{The supernemo project},
  \href{https://doi.org/10.1134/S1063778806120131}{\emph{Physics of Atomic
  Nuclei} {\bfseries 69} (2006) 2096}.

\bibitem{nvdexcdr2023}
X.-G.~Cao, Y.-L.~Chang, K.~Chen, E.~Ciuffoli, L.-M.~Duan, D.-L.~Fang et~al.,
  \emph{N$\nu$dex-100 conceptual design report},
  \href{https://doi.org/10.1007/s41365-023-01360-7}{\emph{Nuclear Science and
  Techniques} {\bfseries 35} (2023) 3}.

\bibitem{Nygren_2018}
D.~Nygren, B.~Jones, N.~López-March, Y.~Mei, F.~Psihas and J.~Renner,
  \emph{Neutrinoless double beta decay with 82sef6 and direct ion imaging},
  \href{https://doi.org/10.1088/1748-0221/13/03/P03015}{\emph{Journal of
  Instrumentation} {\bfseries 13} (2018) P03015}.

\bibitem{10262350}
L.~Lang, Y.~Hu, Z.~Yu, B.~You, T.~Liang, Z.~He et~al., \emph{Design and
  demonstration of digital readout chain in n$\nu$dex experiment},
  \href{https://doi.org/10.1109/TNS.2023.3319231}{\emph{IEEE Transactions on
  Nuclear Science} {\bfseries 70} (2023) 2499}.

\bibitem{Yang_2024}
Y.~Yang, T.~Liang, C.~Gao, D.~Zhang, K.~Chen, H.~Wang et~al., \emph{Design and
  preliminary test results of the charge sensitive amplifier for gain-less
  charge readout in high-pressure tpc},
  \href{https://doi.org/10.1088/1748-0221/19/03/C03031}{\emph{Journal of
  Instrumentation} {\bfseries 19} (2024) C03031}.

\bibitem{Liang_2024}
T.~Liang, D.~Zhang, H.~Wang, C.~Gao, J.~Liu, X.~Sun et~al., \emph{Performance
  of a novel charge sensor on the ion detection for the development of a
  high-pressure avalancheless ion tpc},
  \href{https://doi.org/10.1088/1748-0221/19/04/C04004}{\emph{Journal of
  Instrumentation} {\bfseries 19} (2024) C04004}.

\bibitem{Phan_2017}
N.~Phan, R.~Lafler, R.~Lauer, E.~Lee, D.~Loomba, J.~Matthews et~al., \emph{The
  novel properties of sf6 for directional dark matter experiments},
  \href{https://doi.org/10.1088/1748-0221/12/02/P02012}{\emph{Journal of
  Instrumentation} {\bfseries 12} (2017) P02012}.

\bibitem{opamanual}
``Opax2111.1-nv/rthznoise, lowpower, precisionoperational amplifiers.''

\end{thebibliography}\endgroup

\end{document}